\documentclass[aps,prb,twocolumn,superscriptaddress,showpacs,showkeys]{revtex4-2}

\usepackage{color}
\usepackage{graphicx}
\usepackage{dcolumn}
\usepackage{bm}
\usepackage{float}
\usepackage[mathlines]{lineno}
\usepackage{tabularx}
\usepackage[colorlinks,linkcolor=blue,anchorcolor=blue,citecolor=blue,urlcolor=blue,hyperindex,CJKbookmarks]{hyperref}
\usepackage{footnote}
\usepackage{amsmath}
\usepackage{txfonts}
\usepackage{mathptmx}
\usepackage{ragged2e}
\usepackage{booktabs,makecell, multirow, tabularx}
\usepackage{multirow}
\usepackage{braket}
\usepackage{gensymb}
\usepackage{array}
\usepackage{makecell} 
\usepackage{booktabs} 
\usepackage{svg}
\usepackage{siunitx}

\newcolumntype{L}[1]{>{\raggedright\let\newline\\\arraybackslash\hspace{0pt}}m{#1}}
\newcolumntype{C}[1]{>{\centering\let\newline\\\arraybackslash\hspace{0pt}}m{#1}}
\newcolumntype{R}[1]{>{\raggedleft\let\newline\\\arraybackslash\hspace{0pt}}m{#1}}

\begin{document}

\title{Absence of two-orbital superconductivity in cuprate family: A DFT+DMFT perspective}

\author{Jian-Hong She}
\thanks{These authors contributed equally to this work.}
\affiliation{School of Physics and Beijing Key Laboratory of Opto-electronic Functional Materials $\&$ Micro-nano Devices, Renmin University of China, Beijing 100872, China}\affiliation{Key Laboratory of Quantum State Construction and Manipulation (Ministry of Education), Renmin University of China, Beijing 100872, China}
\author{Jing-Xuan Wang}
\thanks{These authors contributed equally to this work.}
\affiliation{School of Physics and Beijing Key Laboratory of Opto-electronic Functional Materials $\&$ Micro-nano Devices, Renmin University of China, Beijing 100872, China}\affiliation{Key Laboratory of Quantum State Construction and Manipulation (Ministry of Education), Renmin University of China, Beijing 100872, China}

\author{Rong-Qiang He}\email{rqhe@ruc.edu.cn}\affiliation{School of Physics and Beijing Key Laboratory of Opto-electronic Functional Materials $\&$ Micro-nano Devices, Renmin University of China, Beijing 100872, China}\affiliation{Key Laboratory of Quantum State Construction and Manipulation (Ministry of Education), Renmin University of China, Beijing 100872, China}
\author{Zhong-Yi Lu}\email{zlu@ruc.edu.cn}\affiliation{School of Physics and Beijing Key Laboratory of Opto-electronic Functional Materials $\&$ Micro-nano Devices, Renmin University of China, Beijing 100872, China}\affiliation{Key Laboratory of Quantum State Construction and Manipulation (Ministry of Education), Renmin University of China, Beijing 100872, China}\affiliation{Hefei National Laboratory, Hefei 230088, China}

\date{\today}

\begin{abstract}
The recent discovery of high-temperature superconductivity in the bilayer nickelate La$_3$Ni$_2$O$_7$ has spurred intense interest in exploring analogous mechanisms in other transition metal oxides. This raises a pivotal question: can cuprates, as neighbors to nickelates in the periodic table, host similar two-orbital superconductivity? Here, we systematically investigate the electronic structure of a series of designed Ruddlesden-Popper cuprates. Our calculations reveal that the parent compound La$_3$Cu$_2$O$_7$ is a weakly correlated metal, and hole-doping fails to induce strong correlation. We find that the actual valence of the copper cations becomes strikingly pinned around +2.3, far away from the targeted $d^8$ configuration. This valence pinning is attributed to the inherent charge-transfer nature of cuprates. We propose this mechanism as a general principle explaining the robust single-orbital physics consistently observed in the cuprate family, holding true even in materials like the high-$T_c$ superconductor Ba$_2$CuO$_{3+\delta}$ that appear structurally primed for two-orbital activity. Our results therefore conclude that the route towards two-orbital superconductivity is fundamentally obstructed in cuprates, providing a crucial constraint for the future design of high-temperature superconductors.

\end{abstract}


\maketitle

\section{Introduction}
The recent experimental discovery of high-temperature superconductivity in the bilayer nickelate La$_{3}$Ni$_{2}$O$_{7}$, with a transition temperature ($T_c$) reaching 80 K under high pressure, has opened a new frontier in condensed matter physics \cite{Sun-nature,Wang2024_nature,Zhang2024_naturephys,Zhu2024_nature}. A broad consensus has emerged from numerous theoretical studies that this phenomenon is driven by a two-orbital superconducting mechanism, originating from the strong electronic correlation between the Ni $d_{z^2}$ and $d_{x^2-y^2}$ orbitals which are both near a half-filled state \cite{PhysRevB.109.165140,Ouyang2024_prb,Yang2024_natcommun,Liu2023_prl,Lu2024_prl}. A key aspect of this physics is the crucial role of Hund's coupling \cite{Ouyang2024_prb,PhysRevB.109.165140,PhysRevB.109.165154,PhysRevB.110.L235119}. While the bare Hubbard interaction $U$ on Ni-3$d$ orbitals is considered only moderately strong, the substantial Hund's coupling $J_{H}$, which favors high-spin configurations for the $e_g$ electrons, effectively suppressing charge fluctuations and driving the system into a``Hund's metal" regime, significantly enhances electronic correlation \cite{PhysRevB.109.165154,PhysRevB.110.L235119,Shilenko2023_prb,Ouyang2024_prb}. This paradigm, rooted in multi-orbital Hund's physics, is distinct from that of conventional cuprates, where a larger Hubbard $U$ is the primary driver of correlation \cite{Karp2021_prb,Anderson1987_science,Kang2023_npjqm}, and suggests that engineering such multi-orbital effects is a promising route towards higher $T_{c}$.

This breakthrough in nickelates naturally raises a compelling question: can cuprates, the elemental neighbor to nickelates, in the periodic table, be coaxed into a similar two-orbital superconducting state? The Ruddlesden-Popper (RP) phases of cuprates, La$_{n+1}$Cu$_{n}$O$_{3n+1}$, are isostructural to their nickelate counterparts, suggesting the possibility of analogous physics. For instance, the bilayer La$_{3}$Cu$_{2}$O$_{7}$, if synthesized, would be a direct structural analogue of the superconducting La$_{3}$Ni$_{2}$O$_{7}$. The central hypothesis to be tested is whether the Cu-$e_g$ orbitals in such a structure can be tuned to a strongly correlated regime, similar to the Ni-$e_g$ orbitals. This prospect, however, stands in stark contrast to the well-established physics of known high-$T_{c}$ cuprates, whose parent compounds are typically described as single-band Mott or charge-transfer insulators dominated by the Cu $d_{x^2-y^2}$ orbital \cite{Vaknin1987_prl,Keimer2015_nature}.

Realizing a nickelate-like electronic structure in cuprates presents a formidable challenge. In La$_{3}$Ni$_{2}$O$_{7}$, the nominal Ni valence of $+$2.5 results in a 3$d^{7.5}$ configuration, which through strong Hund's coupling, effectively behaves as a strongly correlated $d^{8}$ system--the key ingredient for its superconductivity. To replicate this, copper would need to be stabilized in a highly unusual $d^{8}$ or a proximate configuration (such as $d^{8.2}$). To rigorously test the viability of this route, we propose a systematic strategy of aggressive hole-doping, both through chemical substitution (e.g., creating LaSr$_{2}$Cu$_{2}$O$_{7}$ and Sr$_{3}$Cu$_{2}$O$_{7}$) , aiming to drive the Cu-$e_g$ occupancy towards the strongly correlated half-filling regime.

In this paper, we systematically investigate the electronic structure of this series of designed RP cuprates using first-principles density functional theory combined with dynamical mean-field theory (DFT+DMFT). We find that the parent compound La$_{3}$Cu$_{2}$O$_{7}$ is a weakly correlated metal, far from the desired state. More importantly, our extensive calculations on doped systems reveal a``valence pinning" phenomenon: despite nominal valences reaching as high as $+$4.0, the actual calculated valence of copper saturates robustly around $+$2.3. We demonstrate that this pinning is an intrinsic consequence of the charge-transfer nature of cuprates, where it is energetically favorable for doped holes to reside on the oxygen ligands rather than on the copper sites. Our findings conclude that this inherent electronic characteristic fundamentally obstructs the path towards achieving two-orbital, nickelate-like superconductivity in cuprates, thereby providing a crucial guiding principle for future materials design in the field of high-temperature superconductivity.

\section{Method}
Our theoretical investigation was conducted in two main stages. The initial structural optimization of all studied compounds was performed using the PBEsol exchange correlation functional \cite{Perdew2008_prl} and the projector augmented-wave (PAW) method as implemented in Vienna ab initio simulation package (\textsc{VASP}) \cite{Kresse1993_prb,Kresse1994_prb,Kresse1996_cms,Kresse1996_prb,Kresse1999_prb}. A plane-wave cutoff energy of 520~eV was applied for the Kohn-Sham orbitals, and spin-orbit coupling was not included in our calculations. The initial structure for the parent compound, La$_3$Cu$_2$O$_7$, was obtained by substituting Ni with Cu in the experimental La$_3$Ni$_2$O$_7$ structure. For the doped systems, specific La sites were substituted by Sr, Ca, or Ba in a manner that preserves the overall $I4/mmm$ crystal symmetry. A full structural optimization under fixed target pressure, relaxing both the lattice parameters and all internal atomic positions, was performed for each system until the Hellmann-Feynman force on each atom was less than 0.01 eV per angstrom.
An appropriately dense $k$-point mesh was used for the Brillouin zone integration during the self-consistent field calculations.

The charge fully self-consistent DFT+DMFT calculations were carried out using the eDMFT software package \cite{Haule2010_prb}. The DFT part of this stage was performed with the \textsc{wien}2k code, which is based on the full-potential linearized augmented plane-wave (FP-LAPW) method \cite{Blaha2020_jcp}. A self-consistency loop between DFT and DMFT was performed to ensure charge redistribution effects being fully captured. Each DFT+DMFT cycle included one DMFT calculation followed by a maximum of 100 DFT iterations to converge the charge density. A good convergence was typically achieved within 50 DFT+DMFT cycles, with the convergence criteria for charge and total energy set to $10^{-7}$~e and $10^{-7}$~Ry, respectively. All calculations were performed in the paramagnetic state at a temperature of 290~K. Within the DMFT framework, the Cu-3$d$ $e_g$ ($d_{z^2}$ and $d_{x^2-y^2}$) orbitals were treated as the correlated subspace. Due to the $I4/mmm$ space group symmetry of the bilayer structure, the two Cu atoms within the primitive unit cell are crystallographically equivalent, simplifying the many-body problem to a single-impurity problem. We used a Hubbard interaction $U = 9.0$~eV and a Hund's exchange $J_\text{H} = 1.0$~eV. The value of $U$ was chosen to be larger than that typically used for nickelates ($U = 5.0$~eV) to reflect the stronger localization and correlation effects expected for Cu-3$d$ orbitals. The density-density form of the Coulomb repulsion was employed. Projectors onto the correlated orbitals were constructed using an energy window from $-10$ to $10$~eV with respect to the Fermi level. The resulting quantum impurity problem was solved using the continuous-time quantum Monte Carlo solver in the hybridization expansion (CT-HYB) formulation \cite{Haule2007_prb}. The self-energy was corrected for double-counting using the exact double-counting scheme \cite{Haule2015_prl}. Finally, the real-frequency self-energy, momentum-resolved spectral function, and other related physical quantities were obtained using the maximum entropy method for analytical continuation \cite{Jarrell1996_physrep}.

\section{Results}
In this section, we present our first-principles calculation results to systematically investigate the electronic structure of the La$_3$Cu$_2$O$_7$ family and elucidate why it fails to host two-orbital superconductivity. We begin with the electronic properties of the parent compound, followed by a detailed analysis of the effects of chemical doping.

\subsection{Electronic Structure of the Parent Compound La$_3$Cu$_2$O$_7$}

\begin{figure}[hb]
\centering
\includegraphics[width=8.6cm]{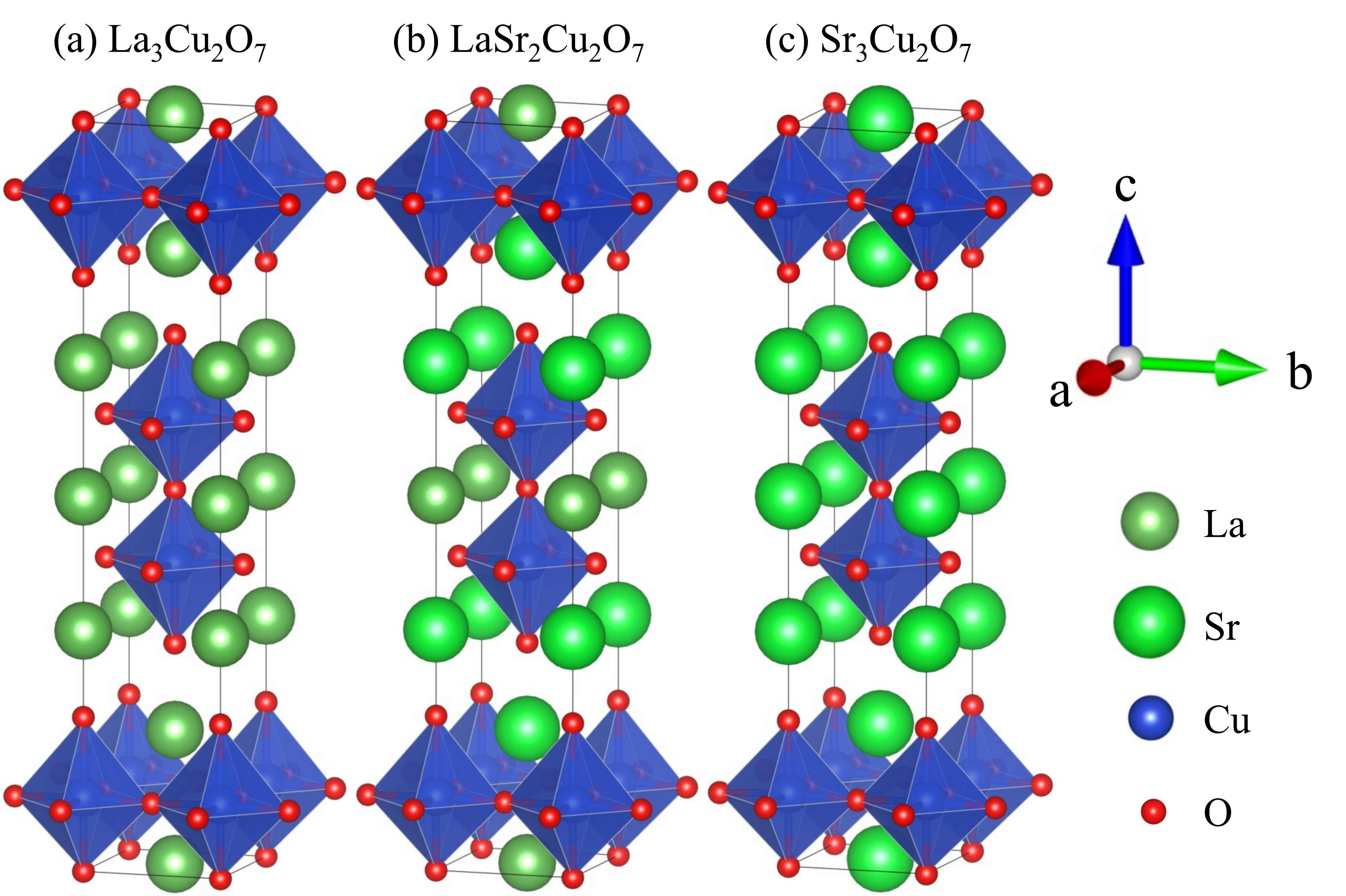}
\caption{The crystal structure of (a) La$_3$Cu$_2$O$_7$, (b) LaSr$_2$Cu$_2$O$_7$ and (c) Sr$_3$Cu$_2$O$_7$ at high pressure (30~GPa).}
\label{fig1}
\end{figure}

 We first investigated the parent compound La$_3$Cu$_2$O$_7$ at a pressure of 30~GPa, inspired by the synthesis conditions of nickelate superconductors. The crystal structure was fully relaxed within DFT, confirming the $I4/mmm$ space group symmetry, consistent with La$_3$Ni$_2$O$_7$, as shown in Fig.~\ref{fig1}. The optimized lattice constants are presented in Table~\ref{tab:lattice_constants_combined}, which also lists the structure for all other systems studied in this work.

\begin{table}[htbp]
    \centering
    \caption{Lattice parameters $a=b$ and $c$ (in angstrom) for La$_3$Ni$_2$O$_7$ and computed RP cuprates under high pressure (30 GPa).}
    \label{tab:lattice_constants_combined}
    \begin{tabular}{lcc}
        \hline\hline 
        & $a=b$  & $c$  \\ 
        \hline 
        La$_3$Ni$_2$O$_7$      & 3.680 & 19.362 \\ 
        La$_3$Cu$_2$O$_7$      & 3.704 & 19.501 \\
        LaSr$_2$Cu$_2$O$_7$    & 3.748 & 19.763 \\
        LaCa$_2$Cu$_2$O$_7$    & 3.634 & 18.623 \\
        LaBa$_2$Cu$_2$O$_7$    & 3.756 & 19.745 \\
        Sr$_3$Cu$_2$O$_7$      & 3.681 & 19.001 \\
        \hline\hline 
    \end{tabular}
\end{table}

\begin{figure*}[htbp]
\centering
\hspace*{-0.5cm}
\includegraphics[width=18.5cm]{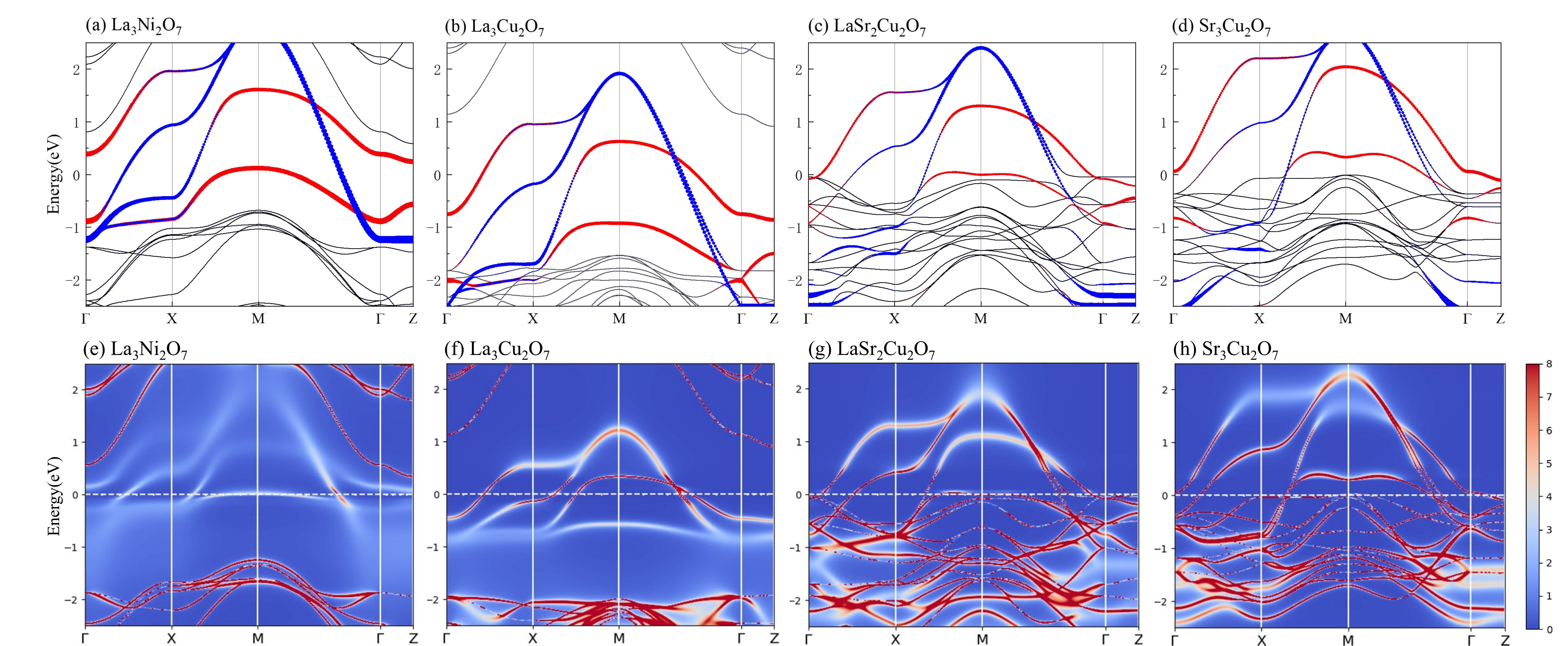}
\caption{Calculated electronic structures of selected RP compounds at 290 K and high pressure. La$_3$Ni$_2$O$_7$ is calculated using the experimental synthesized structure under 29.5 GPa while cuprates are calculated under 30 GPa with DFT relaxed structures.
\textbf{(a-d)} Band structures from DFT calculations for La$_3$Ni$_2$O$_7$, La$_3$Cu$_2$O$_7$, LaSr$_2$Cu$_2$O$_7$, and Sr$_3$Cu$_2$O$_7$. The thicker red and blue lines highlight orbital characters of Ni/Cu $d_{z^2}$ and $d_{x^2-y^2}$, respectively.
\textbf{(e-h)} The corresponding k-resolved spectral functions from DFT+DMFT calculations. The color map denotes the spectral weight.}
\label{fig2}
\end{figure*}

To understand the basic electronic properties, we first calculated the orbital-resolved band structure using DFT, as shown in Fig.~\ref{fig2}. The bands crossing the Fermi level are predominantly of Cu $e_g$ ($d_{z^2}$ and $d_{x^2-y^2}$) character. However, the center of the $e_g$ bands is located significantly above that of La$_3$Ni$_2$O$_7$, indicating that these orbitals are far from the half-filling condition required for strong Mott physics.

\begin{figure}[htbp]
\centering
\includegraphics[width=8.6cm]{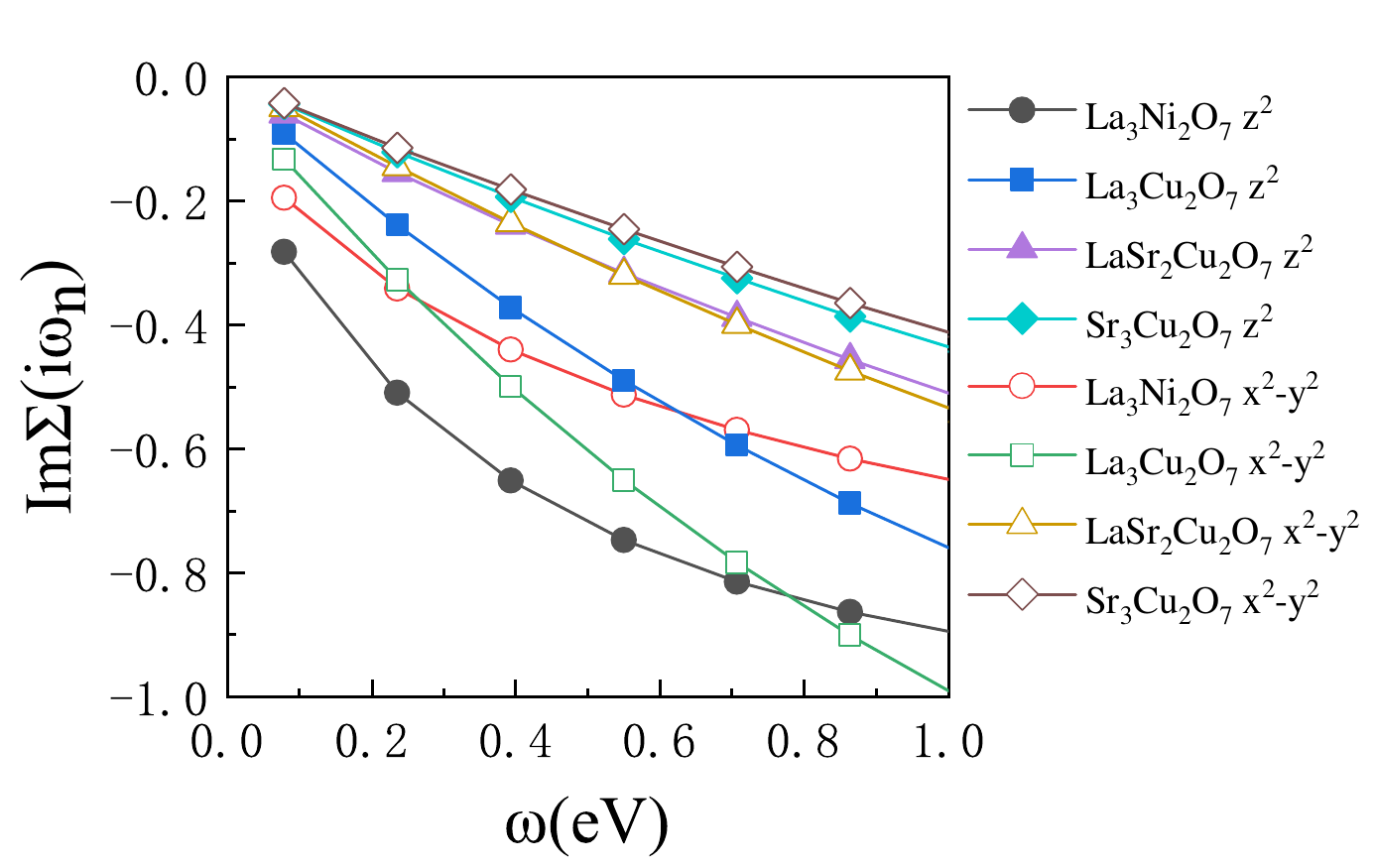}
\caption{Imaginary part of the Matsubara self energy for both $d_{z^2}$ and $d_{x^2-y^2}$ orbitals of Ni/Cu atoms in La$_3$Ni$_2$O$_7$, La$_3$Cu$_2$O$_7$, LaSr$_2$Cu$_2$O$_7$, and Sr$_3$Cu$_2$O$_7$.}
\label{fig3}
\end{figure}

To incorporate local electronic correlation, which is crucial for 3$d$ transition metal compounds, we performed DFT+DMFT calculations. The results are presented in Figs.~\ref{fig2} and~\ref{fig3}, which shows the imaginary part of the orbital-resolved Matsubara self-energy $\text{Im}\Sigma(i\omega_n)$ and the spectral function $A(\omega)$.

Our calculations reveal that La$_3$Cu$_2$O$_7$ is a Fermi liquid. The most telling evidence for this conclusion comes from the low-frequency behavior of $\text{Im}\Sigma(i\omega_n)$ as shown in Fig.~\ref{fig3}. For La$_3$Cu$_2$O$_7$, $\text{Im}\Sigma(i\omega_n)$ exhibits a robust linear dependence as it approaches zero from imaginary axis. This is in stark contrast to the behavior universally observed in the normal state of high-$T_c$ superconductors, which are characterized as ``strange metals'' where $\text{Im}\Sigma(i\omega_n)$ shows a nonlinear dependence on Matsubara frequency (e.g., the case of La$_3$Ni$_2$O$_7$). The Fermi liquid nature of La$_3$Cu$_2$O$_7$ indicates that electron scattering is suppressed at low energies, leading to long-lived quasiparticles. This picture is further corroborated by the spectral function $A(\omega)$ shown in Fig.~\ref{fig2}(b), which displays very sharp and intense quasiparticle peaks at the Fermi level, consistent with the existence of well-defined quasiparticles.

The correlation effect can be quantified using the quasiparticle mass renormalization $m^*/m = Z^{-1}$, where the quasiparticle weight $Z$ is defined as
\begin{equation}
{Z}^{-1} =   1 - \frac{\partial \mbox{Re} \Sigma (\omega)}{\partial \omega} \bigg|_{\omega = 0} .
\label{eq:weight}
\end{equation}
We obtain orbital-differentiated values of $m^*/m = 1.94$ for the $d_{z^2}$ orbital and $m^*/m = 2.24$ for the $d_{x^2-y^2}$ orbital, respectively. These quantitative results, in conjunction with the Fermi-liquid behavior of the self-energy and the sharp quasiparticle peaks discussed above, provide a self-consistent picture. They confirm that La$_3$Cu$_2$O$_7$ is a weakly correlated metal, indicating that the existing electronic correlation is insufficient to drive the Fermi liquid state into the ``strange metal'' regime, which is considered the necessary ground for high-temperature superconductivity.

\subsection{Doping effects in LaM$_2$Cu$_2$O$_7$ (M=Sr, Ca, Ba)}
Given that the parent compound is weakly correlated due to its $e_g$ orbitals being far from half filling, a natural strategy is to introduce hole doping to oxidize Cu$^{2+}$ towards Cu$^{3+}$ ($d^8$). We attempted this by substituting trivalent La$^{3+}$ with divalent cations M$^{2+}$ (M = Sr, Ca, Ba).

We first focus on the LaSr$_2$Cu$_2$O$_7$ system. As expected, our DFT calculations (Fig.~\ref{fig2}(c)) show that hole doping leads to a rigid upward shift of the Fermi level by approximately 0.4~eV compared to the parent compound. This simple band-filling picture suggests a significant reduction in the Cu $d$-electron count. However, our DFT+DMFT results reveal a starkly different and more complex reality. The calculated Cu $d$-shell occupancy is 8.844, corresponding to an effective valence of only +2.156, while the undoped La$_3$Cu$_2$O$_7$ has a Cu $d$-shell occupancy 9.029, corresponding to a typical $d^9$ configuration. Despite the nominal introduction of one hole per formula unit, the Cu ion is only marginally oxidized, remaining very close to a $d^9$ configuration and far from the $d^{8.2}$ state in La$_3$Ni$_2$O$_7$ \cite{Ouyang2024_prb}.

Counter-intuitively, this shift in occupancy towards the half-filled $d^8$ configuration does not lead to stronger electronic correlation. In fact, the system becomes less correlated upon doping. This is quantitatively demonstrated by the quasiparticle mass renormalization as shown in Table~\ref{tab:orbital_data_updated_simple}, which decreases relative to the parent compound, indicating a reduction in the correlation effects.

This behavior is characteristic of transition metal oxides described by the Zaanen-Sawatzky-Allen (ZSA) scheme \cite{Zaanen1985_prl}. In this framework, the electronic properties are governed by the competition between the on-site $d$-$d$ Coulomb repulsion $U$ and the $p$-$d$ charge-transfer energy $\Delta$. Our results strongly suggest that the La$_3$Cu$_2$O$_7$ system is of the charge-transfer type ($\Delta < U$), where the energy cost to create a hole on the oxygen ligand's $p$ orbitals is lower than creating a $d^8$ site on copper. This directly explains why hole doping primarily creates holes on the O-$p$ orbitals, leaving the Cu valence largely unchanged.

Furthermore, this charge-transfer nature has a more subtle consequence on the correlation strength itself. The primary effect of populating the O-$p$ orbitals significantly enhance the $p$-$d$ hybridization. This is directly reflected in our calculated Cu $e_g$ spectral function (Fig.~\ref{fig2}(g)), which reveals a noticeable broadening of the overall $d$-orbital bandwidth in LaSr$_2$Cu$_2$O$_7$ compared to the parent compound. This increased bandwidth is a clear signature of enhanced electron itinerancy and reduced localization. From a many-body perspective, the enhanced $p$-$d$ hybridization constitutes an efficient dynamical screening mechanism. The rapid charge fluctuations between the newly introduced, itinerant oxygen $p$-holes and the localized copper $d$-orbitals effectively screen the on-site Coulomb repulsion $U$, leading to a substantially reduced effective interaction $U_\text{eff}$ on the copper sites and a weaker overall correlation.

\begin{figure*}[htbp]
\centering
\hspace*{-0.5cm}
\includegraphics[width=18.5cm]{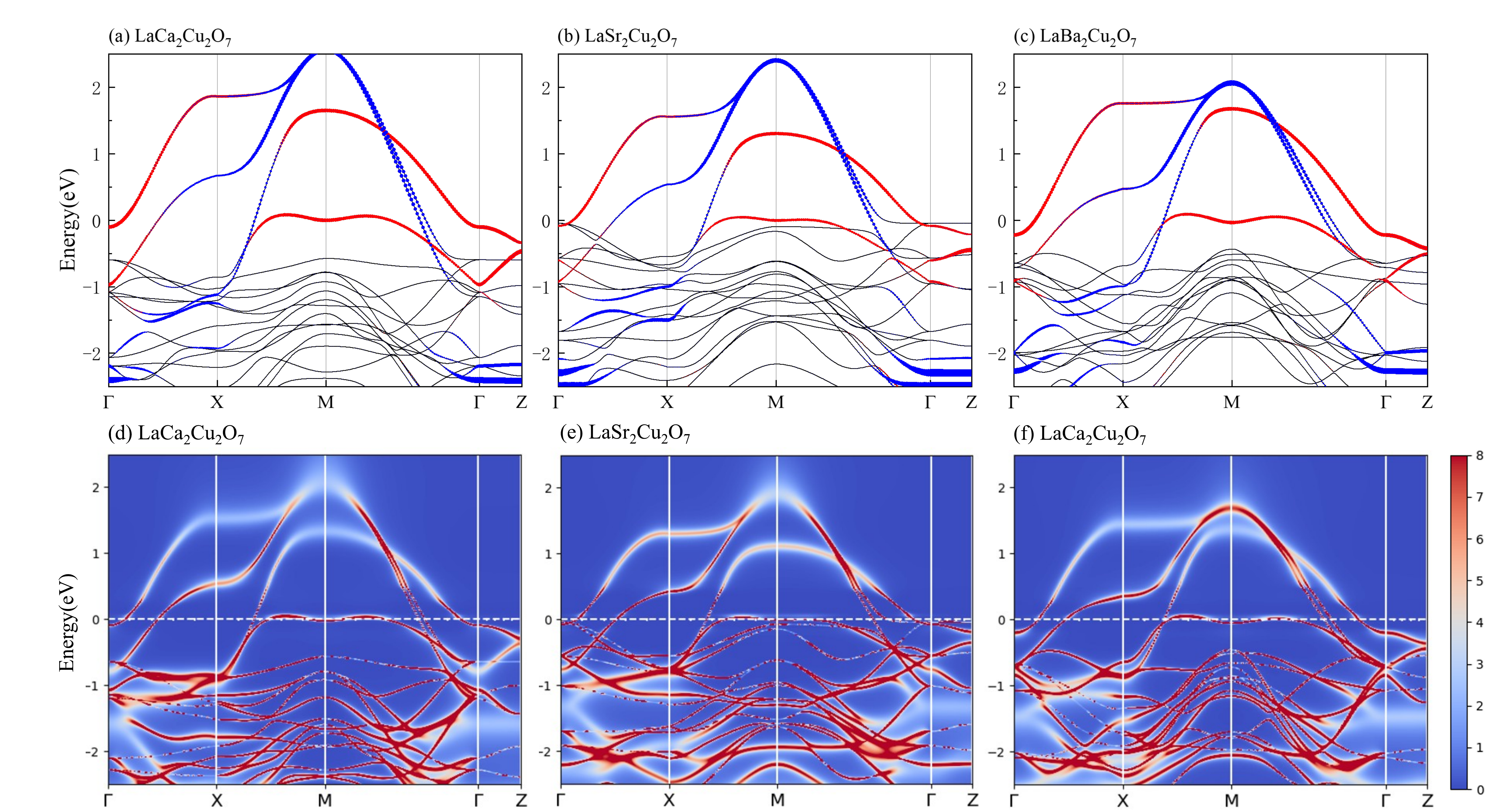}
\caption{Calculated electronic structures of selected RP compounds at 290K and 30 GPa.
\textbf{(a-c)} Band structures from DFT calculations for LaSr$_2$Cu$_2$O$_7$, LaCa$_2$Cu$_2$O$_7$ and LaBa$_2$Cu$_2$O$_7$. The thicker red and blue lines highlight orbital characters of Ni/Cu $d_{z^2}$ and $d_{x^2-y^2}$, respectively.
\textbf{(d-f)} The corresponding k-resolved spectral functions from DFT+DMFT calculations. The color map denotes the spectral weight.}
\label{fig4}
\end{figure*}

To solidify this conclusion and rule out any effects specific to the dopant ion, we also performed calculations for M = Ca and Ba. Fig.~\ref{fig4} presents a direct comparison of the Cu $e_g$ spectral functions for all three dopants (Sr, Ca, Ba). The spectra are similar, demonstrating that the electronic state of copper is insensitive to the choice of the divalent dopant, despite their different electronegativities. This is further corroborated by the calculated Cu $d$-shell occupancies listed in Table~II, which show only minor variations across the three systems. This collective evidence robustly supports our conclusion that the intrinsic charge-transfer nature of the Cu-O bonds in this structure prevents the effective hole doping of the Cu $e_g$ orbitals.

\subsection{Effects of Further Doping}
Finally, we explored a further avenue to potentially drive the system into a more strongly correlated state: increasing the doping concentration to its limit.

We considered the case of extreme hole doping leading to the chemical formula Sr$_3$Cu$_2$O$_7$. Nominally, this corresponds to a Cu valence of +4, which should place the $e_g$ occupancy well below half-filling. Once again, our DFT+DMFT calculations show that the system resists this high oxidation state. The calculated Cu valence is only +2.3, and as shown in Fig.~\ref{fig2}(h), the system remains a weakly-correlated metal with sharp quasiparticle peaks. This demonstrates that even under aggressive doping, the holes continue to localize on the O-$p$ orbitals. Key quantitative metrics, including the Cu $d$-shell occupancy and the quasiparticle weight $Z$, are compiled in Table~\ref{tab:orbital_data_updated_simple}. These values show negligible changes, indicating that even extreme hole doping is ineffective to tune the $U/\Delta$ ratio or hopping parameters in a way that would enhance correlation or promote charge transfer to the Cu sites.

\begin{table}[h!]
\centering
\caption{
    Local orbital occupation number $N_d$, total $d$-orbital occupation number and effective mass enhancement $m^*/m$ of the Ni and Cu atoms in several systems. The $t_{2g}$ orbitals are assumed to be fully occupied as they lie far below the Fermi level.
}
\label{tab:orbital_data_updated_simple}
\begin{tabular}{@{}lccccc@{}}
\toprule
 & \multicolumn{3}{c}{$N_d$} & \multicolumn{2}{c}{$m^*/m$} \\
\cmidrule(lr){2-4} \cmidrule(lr){5-6}
 & $d_{z^2}$ & $d_{x^2-y^2}$ & total & $d_{z^2}$ & $d_{x^2-y^2}$ \\
\midrule
Ni in La$_3$Ni$_2$O$_7$ & 1.136 & 1.056 & 8.192 & 3.73 & 2.52 \\
Cu in La$_3$Cu$_2$O$_7$ & 1.604 & 1.425 & 9.029 & 1.94 & 2.24 \\
Cu in LaSr$_2$Cu$_2$O$_7$ & 1.453 & 1.391 & 8.844 & 1.60 & 1.61 \\
Cu in LaCa$_2$Cu$_2$O$_7$ & 1.368 & 1.401 & 8.769 & 1.62 & 1.49 \\
Cu in LaBa$_2$Cu$_2$O$_7$ & 1.375 & 1.448 & 8.823 & 1.63 & 1.53 \\
Cu in Sr$_3$Cu$_2$O$_7$ & 1.329 & 1.412 & 8.741 & 1.49 & 1.46 \\
\bottomrule
\end{tabular}
\end{table}

\section{Discussion}
In this work, we have systematically investigated the viability of inducing nickelate-like two-orbital superconductivity in a series of designed RP cuprates using DFT+DMFT. Our calculations show that the parent compound La$_3$Cu$_2$O$_7$ is a weakly-correlated metal, with its Cu-$e_g$ orbitals far from the half-filled state required for strong Hund's physics. More critically, our extensive study of doped systems reveals that hole-doping strategies fail to drive the system into a strongly correlated, two-orbital active regime.

Therefore, we conclude that the route towards nickelate-like two-orbital superconductivity is fundamentally obstructed in these RP-type cuprates. This raises a crucial question: is this obstruction specific to the RP-phase structure, or does it stem from a more universal principle governing all cuprates? To address this, it is instructive to examine other cuprate systems that, on the surface, appear to be promising candidates for two-orbital physics.

A compelling case is the recently synthesized high-temperature superconductor Ba$_2$CuO$_{3+\delta}$ ($\delta \approx 0.25$), which exhibits a $T_c$ of 73~K despite being in a heavily hole-overdoped regime~\cite{Li2019_pnas}. Structurally, it possesses the shortest apical oxygen distance among known hole-doped cuprates, leading to a compressed octahedral environment along $z$-axis, raising the energy of the $d_{z^2}$ orbital above the $d_{x^2-y^2}$ orbital, seemingly creating the perfect condition for two-orbital activity at the Fermi level. However, detailed DFT+DMFT investigations have revealed a starkly different reality~\cite{wang2025correlated}. The key lies in the detailed analysis of the three inequivalent copper sites. For the six-coordinate Cu(o) atom, a naive analysis in a global coordinate system suggests a compressed octahedron, which should elevate the $d_{z^2}$ orbital energy and promote a two-orbital state. However, the theoretical work demonstrated that by adopting a local coordinate system aligned with the true elongated axis of the Cu(o)O$_6$ octahedron, the electronic structure reverts to the canonical cuprate picture: a fully occupied, energetically deep $d_{z^2}$-like orbital and a single, partially-filled $d_{x^2-y^2}$-like orbital around the Fermi level. This active orbital is strongly correlated, evidenced by a large mass enhancement ($m^*/m \approx 2.47$), confirming its role as a correlated metal. Furthermore, the analysis of the four-coordinate copper atoms reinforces this single-orbital conclusion. The Cu(p) atom, forming the backbone of the quasi-one-dimensional chains, also settles into a single-orbital state with a strongly correlated $d_{x^2-y^2}$-like orbital ($m^*/m \approx 2.22$) after a similar local coordinate transformation. In contrast, the third site, the bridging Cu(b) atom, is shown to be a weakly correlated band insulator, rendering it electronically inert and largely irrelevant to the superconductivity. Crucially, the authors argue that the system's high-$T_c$ superconductivity originates specifically from the strongly correlated, single-orbital physics within the quasi-two-dimensional electronic network of the Cu(o) planes.

Taken together, our findings on RP-cuprates and the evidence from Ba$_2$CuO$_{3+\delta}$ point to a unified picture: the inherent electronic nature of the Cu-O bond prevents the realization of a stable, nickelate-like two-orbital active state in cuprates. The underlying reason is the strong Cu-O hybridization and the high energy cost of a Cu $d^8$ configuration relative to $d^9$. Even under extreme structural conditions, such as the compressed octahedra in Ba$_2$CuO$_{3+\delta}$, or heavy doping as explored in our work on La$_3$Cu$_2$O$_7$, the system preferentially places holes on the oxygen ligands rather than creating a second hole on the copper ion. This robust charge-transfer characteristic ensures the persistence of a single-orbital $d_{x^2-y^2}$ picture.

Therefore, we conclude that the route towards two-orbital superconductivity via the mechanism found in La$_3$Ni$_2$O$_7$ is fundamentally obstructed across the cuprate family. This obstruction is not a matter of structure or insufficient doping, but is rooted in the inherent charge-transfer properties of copper in an oxide environment. Our findings provide a crucial guiding principle for the future rational design of high-temperature superconductors.

\begin{acknowledgments}
This work was supported by the National Key R\&D Program of China (Grants No. 2024YFA1408601 and No. 2024YFA1408602) and the National Natural Science Foundation of China (Grant No. 12434009). J.X.W. was also supported by the Outstanding Innovative Talents Cultivation Funded Programs 2025 of Renmin University of China. Z.Y.L. was also supported by the Innovation Program for Quantum Science and Technology (Grant No. 2021ZD0302402). Computational resources were provided by the Physical Laboratory of High Performance Computing in Renmin University of China.
\end{acknowledgments}

\newpage
\bibliography {cu327}

\end{document}